\documentclass[preprint, showpacs,preprintnumbers,amsmath,amssymb,nofootinbib]{revtex4}
\usepackage{mathrsfs}
\usepackage{amsmath}
\usepackage[colorlinks, linkcolor=blue, citecolor=red, urlcolor=red]{hyperref}
\usepackage{amsfonts}
\usepackage{latexsym}
\usepackage{amsfonts}
\usepackage{graphicx}
\usepackage{epsf}
\usepackage{dcolumn}
\usepackage{bm}
\newcommand{\bea}[1]{\begin{eqnarray}\label{#1}}
 \newcommand{\eea}{\end{eqnarray}}

 \def\gsim{ \lower .75ex \hbox{$\sim$} \llap{\raise .27ex \hbox{$>$}} }
 \def\lsim{ \lower .75ex \hbox{$\sim$} \llap{\raise .27ex \hbox{$<$}} }
\def\/{\over}

\begin{document}

\title{\bf Emergent universe in spatially flat cosmological model}

\author{ Kaituo Zhang$^1$, Puxun Wu$^{2,3}$, Hongwei Yu$^{1,2,3,}$\footnote{Corresponding
author: hwyu@hunnu.edu.cn}}
\address{ $^1$Department of Physics and Key Laboratory of Low Dimensional
Quantum Structures and Quantum Control of Ministry of Education,
Hunan Normal University, Changsha, Hunan 410081, China\\
$^2$Center for Nonlinear Science and Department of Physics, Ningbo
University,  Ningbo, Zhejiang 315211, China\\
$^3$Kavli Institute for Theoretical Physics China, CAS, Beijing 100190, China}

\begin{abstract}
The scenario of an emergent universe  provides a promising resolution to the big bang singularity in universes with positive or negative spatial curvature. It however remains unclear whether the scenario can be successfully implemented in a spatially flat universe which seems to be favored by present cosmological observations.
In this paper, we study the stability of Einstein static state solutions in  a spatially flat  Shtanov-Sahni  braneworld scenario.  With a negative dark radiation term included and assuming a scalar field as the only matter energy component, we find that the universe can stay at an Einstein static state past  eternally and then  evolve to an inflation phase naturally as the scalar field climbs up its potential slowly.  In addition, we also propose a concrete potential of the scalar field that realizes this  scenario.
\end{abstract}

\pacs{98.80.Cq, 04.50.Kd}

\maketitle
\section{Introduction}
Although most of the mysteries in the standard hot big bang cosmological model can be  resolved by  an inflation epoch~\cite{Starobinsky1980,Guth1981},  the existence of the big bang singularity  in the very early universe remains an open issue. To resolve  the problem, Ellis {\it et al.}  proposed an emergent  scenario~\cite{Ellis20041,Ellis20042},   in which  the universe originates from an  Einstein static state rather than a big bang singularity with the assumption that the spatial curvature is positive, stays there past eternally and then evolves  to an inflation.
However,  the original version of the emergent scenario  in the framework of  general relativity does not seem to resolve the big bang singularity problem successfully as expected since the Einstein static state is unstable under homogeneous perturbations~\cite{Eddington}\footnote {Let us note here that the Einstein static universe has been shown however, by Gibbons~\cite{Gibbons} and Barrow et al.~\cite{Barrow0}, to be rather stable against
inhomogeneous distortions.}.  On the other hand,  in the  very early era, the standard Friedmann equation that governs the dynamics of the universe is very likely to be modified, since  the universe is presumably under extreme physical conditions. In this regard,  it has been found that, in theories including braneworld scenario, modified gravity, quantum gravity, and so on,  a stable Einstein static state universe  can be obtained~\cite{Gergely,Barrow0,Barrow, Canonico, Mulryne, Wu, Parisi}. So,   an emergent universe can still  be regarded as a viable resolution of the big bang singularity in those cases.

Furthermore, a stable Einstein static state  can also exist in an open  universe in Horava-Lifshitz gravity, $f(T)$ gravity, massive gravity and  loop quantum cosmology with modifications to the gravitational sector~\cite{Parisi, Wu, Canonico}  besides the aforementioned  positive curvature cosmological models. Thus, the big bang singularity problem can be resolved  successfully  in the emergent universe scenario irrespective of whether the spatial curvature is positive or negative.

However, the astronomical observations indicate that the universe is very likely spatially flat~\cite{Komatsu2011}.
Therefore, it remains interesting to examine whether the emergent scenario  can be  successfully implemented in a spatially flat universe.  In this paper, we plan to  research the stability of Einstein static state universe in the Shtanov-Sahni  (SS) braneworld scenario without the spatial curvature term~\cite{Shtanov2003}.
 The  braneworld theory is based on superstring theory (M theory),  in which our $1+3$ dimension observable universe (``brane") is  embedded in a $1+3+d$ dimension spacetime (``bulk"). Ordinary particles and fields are confined on the brane while gravity has no such constraint and can access the bulk freely. The SS brane scenario is obtained when the bulk  only has a noncompact timelike fifth dimension and the bane tension is negative. If this extra dimension is spacelike and the brane tension is positive, one then obtains the famous  Randall-Sundrum (RS) braneworld~\cite{Randall19991,Randall19992}.

 Let us note that the extra dimensions are usually assumed to be spacelike. However, the existence of extra timelike dimensions is a possibility that one can not excluded a prior.
 The main problems for typical theories with extra timelike dimensions are that they suffer from pathologies such as  unitarity %probability
 and causality violations~\cite{Yndurain}. These problems  arise from the fact that the Kaluza-Klein (KK) gravitons have an imaginary mass and behave like tachyons, which lead to an imaginary part in the effective gravitational potential of two objects. The violation of causality and unitarity %probability
  in the low-energy processes occur if the induced imaginary part is interpreted as an amplitude for the disappearance into nothing.  Interpreted this way, rather severe bounds  can be put on the size of extra timelike dimension by the experiments.  Interestingly, the induced imaginary part of the gravitational potential can also be interpreted as an artifact of the fictitious decay into the unphysical negative energy tachyons, and then the size of extra timelike dimension can be within the reach of the proposed gravitational experiments~\cite{Dvali}. Nevertheless, extra timelike  dimensions have been considered for the purpose of addressing the cosmological constant problem in KK theories~\cite{Ya},  reconciling a solution of the hierarchy problem with the cosmological expansion of the universe~\cite{Chaichian} and obtaining the  bounce cosmic scenario~\cite{Shtanov2003, Maier}. At the same time, it has been demonstrated that  there are no tachyons or ghosts in the five-dimensional Einstein-Hilbert-Gauss-Bonnet theory with extra timelike dimension~\cite{Iglesias},
  and  the appearance of massless ghosts in an effective four-dimensional theory can be avoided by considering topological criteria in KK theories with extra compactified timelike dimensions~\cite{Ya2}. Furthermore, the avoidance of propagating tachyonic states may also be achieved in the case of
a noncompact timelike extra dimension~\cite{Iglesias} as is in the SS braneworld scenario we are considering  in this paper.

 \section{ Friedmann equation in the SS braneworld}

 The action for the SS braneworld scenario can be written in the following general form~\cite{Shtanov2003}
\begin{eqnarray}
S&=&M^3\bigg[\int_{bulk}\sqrt{-\epsilon g}d^5x(\mathcal{R}-2\Lambda)-2\epsilon\int_{brane}\sqrt{-h}d^4xK\bigg]\\ \nonumber &&+\int_{brane}\sqrt{-h}d^4x(m^2R-2\sigma)
+\int_{brane} \sqrt{-h}d^4xL(h_{ab},\phi)\;,
\end{eqnarray}
where $\mathcal{R}$ is the scalar curvature of the five-dimensional metric $g_{ab}$ in the bulk, and $R$ is the scalar curvature of the induced metric $h_{ab} = g_{ab} - \epsilon n_an_b$ on the brane with $n_a$ being the vector field of the inner unit normal to the brane. $K$  is the trace of the symmetric tensor of the extrinsic curvature $K_{ab} = h^c_a\triangledown_cn_b$ of the brane. $\epsilon=1$ and $-1$ correspond to  the spacelike and timelike extra dimensions, respectively.  $L(h_{ab},\phi)$ denotes the Lagrangian density of the four-dimensional matter fields $\phi$ restricted on the brane. $M$ and $m$ are  respectively the five-dimensional and four-dimensional Planck masses, $\Lambda$ is the five-dimensional cosmological constant and $\sigma$ is the brane tension. $g$ and $h$ are, respectively, the determinants of the matrix of the metric in the bulk and on the brane.  For the case of an arbitrary signature of the extra dimension and an arbitrary Ricci-flat vacuum brane, the five-dimension metric has the form
\begin{eqnarray} d s^2=\epsilon dy^2+\exp\left(-\frac{\epsilon \sigma}{3M^3}y\right)h_{\alpha\beta}dx^\alpha dx^\beta\;.\end{eqnarray}
Here the brane is situated at $y=0$  and the bulk coordinate is in the range $y\geq 0$.

For a homogeneous and isotropic spatially-flat universe on the brane, whose metric  is expressed as $h_{\alpha\beta}=diag(-1, a(t), a(t), a(t))$ with $a$ being the scale factor and $t$ the cosmic time,  the standard Friedmann equation  is found to be modified  as
\begin{eqnarray}
H^2= \frac{2\epsilon\sigma}{M^6}\rho+\frac{\epsilon}{M^6}\rho^2+\frac{C}{a^4}+\frac{1}{3}\bigg(\frac{\Lambda}{2}+\frac{\epsilon\sigma^2}{3M^6}\bigg)\;,
\end{eqnarray}
where $H=\frac{\dot{a}}{a}$ with an overdot denoting a derivative with respect to $t$,  $\rho$ is the cosmic energy density,  and $C$ is an integration constant characterizing  an effective ``dark radiation", which is
induced from the projection of the five-dimensional Weyl tensor on the brane when the bulk space is not conformally flat~\cite{Shiromizu, Mukohyama, Ida}. $\epsilon=1$ and $\sigma>0$ correspond to the RS brane model and  a conformally flat bulk corresponds to $C=0$.  Redefining
\begin{eqnarray}
G_N=\frac{3\epsilon\sigma}{4\pi M^6}\;,\quad
\rho_c=-2\sigma\;,\quad
\Lambda_{eff}=\frac{\Lambda}{2}+\frac{\epsilon\sigma^2}{3M^6}\;,
\end{eqnarray}
and ignoring the effective cosmological constant on the brane, i.e., letting $\Lambda_{eff}=0$, we have
\begin{eqnarray}\label{FEq}
H^2=\frac{8\pi G_N}{3}\bigg(\rho-\frac{\rho^2}{\rho_c}\bigg)+\frac{C}{a^4}\;.
\end{eqnarray}
If $C=0$,  then $H=0$ when $\rho=\rho_c$. This implies that a collapsing brane world-volume is able to undergo a non-singular bounce~\cite{Shtanov2003}.  When $C=0$, the above corrected Friedmann equation can also be obtained  in loop quantum cosmology~\cite{Ashtekar}.   Note that nonlinear energy density corrections to the Friedmann equation were also studied in Ref.~\cite{Markov}.  Hereafter, the natural unit where $8\pi G_N=1$ will be adopted for simplicity.  The stabilities of Einstein static state solutions in positive curvature SS brane model have been studied in Ref.~\cite{Mulryne}, and a successful emergent universe has been obtained in both the $C=0$ and $C>0$ cases.
Since the present observation favors a flat universe strongly, a zero spatial curvature is considered in this paper.  In order to get  Einstein static state solutions in the spatially flat case,   different  from \cite{Mulryne}, here we assume   a negative dark radiation term  ($C<0$).

By assuming that  a scalar field rolling along its potential $V$ is the only matter energy component  on the brane,  the cosmic energy density and pressure can be expressed as   \begin{eqnarray}\label{e1}
\rho=\frac{1}{2}\dot{\phi}^2+V\;,\qquad
p=\frac{1}{2}\dot{\phi}^2-V\;,
\end{eqnarray}
and the dynamics of the scaler field is governed by
\begin{eqnarray}\label{e2}
\ddot{\phi}+3H\dot{\phi}+V'=0\;.
\end{eqnarray}
Differentiating Eq.~(\ref{FEq}) with respect to cosmic time $t$ and using Eqs.~(\ref{e1}, \ref{e2}), one can obtain
\begin{eqnarray}\label{dotH}
\dot{H}=-\bigg(1-\frac{2\rho}{\rho_c}\bigg)(\rho-V)-2\frac{C}{a^4}\;.
\end{eqnarray}
Combining Eq.~(\ref{FEq}) and Eq.~(\ref{dotH}), we have
\begin{eqnarray}\label{ddota/a}
\frac{\ddot{a}}{a}=H^2+\dot{H}=-5H^2+4\frac{C}{a^4}+V +\rho-2V\frac{\rho}{\rho_c}\;.
\end{eqnarray}
In the next section, a constant potential is considered,  which is a good approximation if the variation of potential is very slow with time.

\section{The Einstein static state solution}
The Einstein static state solution satisfies the conditions $\dot{a}=0$ and $\ddot{a}=0$, which imply
\begin{eqnarray}
a=a_{Es},\;\;\;\;H(a_{Es})=0.
\end{eqnarray}
Since $C<0$, Eq.~(\ref{dotH}) shows that the existence of an Einstein static state solution requires $\rho<\frac{\rho_c}{2}$. From Eqs.~(\ref{FEq}) and~(\ref{dotH}), we find that, in a static state universe,  the cosmic energy density  must satisfy
\begin{eqnarray}\label{rhoes}
\rho_{Es\pm}=\frac{1}{8}\bigg(\rho_c+6V\pm\sqrt{{\rho^2_c}-36\rho_c V+36V^2}\bigg)\;,
\end{eqnarray}
which means that the number of  Einstein static state solutions depends on the value of the potential. When $V<0$, there is only one solution with $\rho_{Es}=\rho_{Es+}$.  When $V=0$, we have $\rho_{Es+}=\frac{1}{4}\rho_c$ and $\rho_{Es-}=0$, and there is also only one critical solution in this case since   $\rho_{Es-}=0$ is ruled out by Eq.~(\ref{FEq}).  For a positive potential $V>0$,  the physical meaningfulness requires ${\rho^2_c}-36\rho_c V+36V^2\geq 0$, which leads to $V\leq\frac{1}{6}(3-2\sqrt{2})\rho_c$ or $V\geq\frac{1}{6}(3+2\sqrt{2})\rho_c$.  However, the existence of static state solutions requires $\rho<\frac{\rho_c}{2}$.  As a result, $V\geq\frac{1}{6}(3+2\sqrt{2})\rho_c$ should be discarded. If $0<V<V_{crit}$ where  $V_{crit}\equiv \frac{1}{6}(3-2\sqrt{2})\rho_c$, there are two Einstein static state solutions:
\begin{eqnarray}
a_{Es}=a_{Es+}, \quad a_{Es}=a_{Es-},\end{eqnarray}
with
\begin{eqnarray}\label{aes}
\frac{1}{a_{Es\pm}^4}=\frac{-1}{32C}\bigg(\rho_c+12V-12\frac{V^2}{\rho_c}\pm \bigg(1-2\frac{V}{\rho_c}\bigg)\sqrt{\rho^2_c-36\rho_c V+36V^2}\bigg).
\end{eqnarray}
Once $V=V_{crit}$, these two solutions coincide with each other and there is only one equilibrium point.
Using Eq.~(\ref{FEq}), we can express $\rho$ in terms of $a$ and $H$
\begin{eqnarray}
\rho=\frac{\rho_c}{2}-\sqrt{\rho_c\bigg(3\frac{C}{a^4}-3H^2+\frac{\rho_c}{4}\bigg)}.
\end{eqnarray}
Consequently, Eq.~(\ref{ddota/a}) becomes
\begin{eqnarray}\label{ddota/arew}
\frac{\ddot{a}}{a}=-5H^2+\frac{4C}{a^4}+\frac{\rho_c}{2}-\bigg(\frac{1}{2}-\frac{V}{\rho_c}\bigg)\sqrt{\rho_c\bigg(12\frac{C}{a^4}-12H^2+\rho_c\bigg)}.
\end{eqnarray}

Now we study the stability of critical points. For convenience, we introduce two variables
\begin{eqnarray}
x_1=a\;,\;\;\;\;x_2=\dot{a}\;.
\end{eqnarray}
It is then easy to obtain the following equations
\begin{eqnarray}
\dot{x_1}=x_2\;,
\end{eqnarray}
\begin{eqnarray}
\dot{x_2}=-5\frac{x_2^2}{x_1}+4\frac{C}{x_1^3}+\frac{1}{2}\rho_c x_1-\bigg(\frac{1}{2}-\frac{V}{\rho_c}\bigg)x_1\sqrt{\rho_c\bigg(12\frac{C}{x_1^4}-12\frac{x_2^2}{x_1^2}+\rho_c\bigg)}\;,
\end{eqnarray}
With these variables, the critical points of static state solutions correspond to  $x_1=a_{Es}$, and $x_2=0$. According to the Lyapunov's method, the stability of a critical point is determined by the eigenvalues of the coefficient matrix resulting from linearizing the system described by the above two equations near the critical point.  After a careful calculation, we obtain the eigenvalue $\lambda^2$
\begin{eqnarray}
\lambda^2=-12\frac{C}{a_{Es}^4}+V+12\frac{C}{a_{Es}^4}\frac{\rho_c-2V}{\rho_c-2\rho_{Es}}+\frac{(\rho_c-2V)\rho_{Es}}{\rho_c}\;.
\end{eqnarray}
If $\lambda^2<0$, the corresponding equilibrium point is a stable center  point,  otherwise it is a saddle one.  Substituting  Eqs.~(\ref{rhoes}) and ~(\ref{aes}) into the above equation, one has
\begin{eqnarray}
\lambda^2_\pm=\frac{\rho_c^3-38\rho_c^2 V+108\rho_cV^2-72V^3\pm (\rho_c^2+12\rho_cV-12V^2)\sqrt{\rho_c^2-36\rho_cV+36V^2}}{2\rho_c(-3\rho_c+6V\pm\sqrt{\rho_c^2-36\rho_cV+36V^2})},
\end{eqnarray}
with  $\lambda_\pm$ corresponding to  $a_{Es\pm}$ or $\rho_{Es\pm}$, respectively.

\begin{figure}[htbp]
\includegraphics[width=7cm]{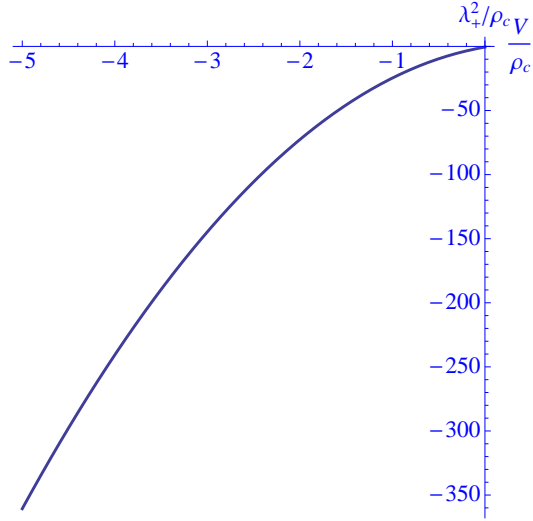}
\caption{\label{f0} The eigenvalue for the case $V\leq0$.   }
\end{figure}

\begin{figure}[htbp]
\includegraphics[width=7cm]{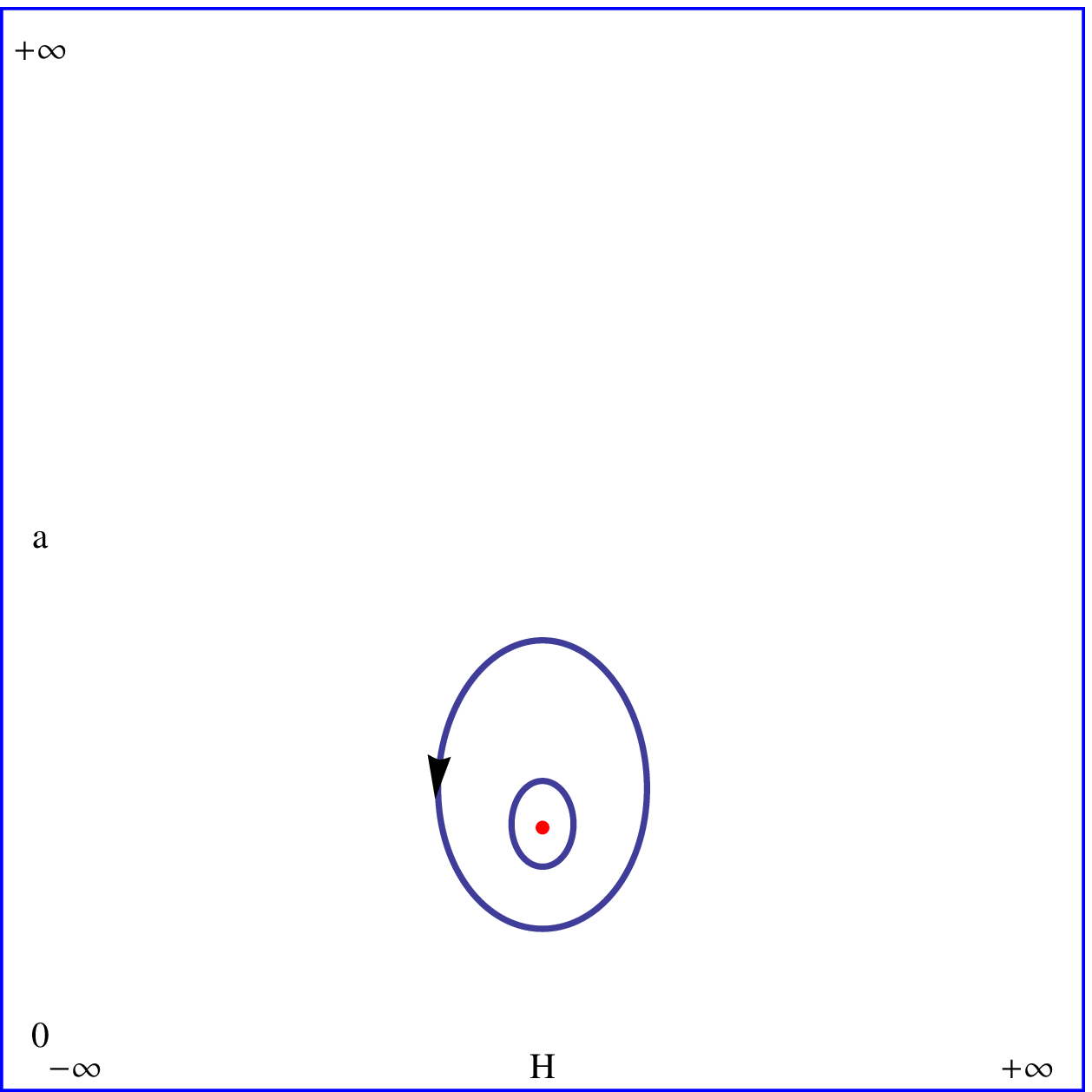}
\caption{\label{f1} The phase portraits in ($a, H$) plane for $V\leq0$.  The parameters are set as $\rho_c=1$, $C=-1$, and $V=-0.01$. There is a stable Einstein static state universe at $a=a_{Es+}$. The red solid point
denotes the stable center point. Circle trajectories correspond to the cases of $a$ differing from  the stable central point initially.  The arrow represents the direction of time evolution. The axes have been compactified using the relations $x(t) = \arctan(H)$ and $y(t) = \arctan(\ln a)$. }
\end{figure}

For the case $V\leq0$, there is only one critical solution $a_{Es}= a_{Es+}$.
Its eigenvalue is $\lambda^2_+$. Fig.~(\ref{f0}) shows that $\lambda^2_+<0$, which implies that this critical point is stable. In Fig.~(\ref{f1}), we plot the phase portraits in ($a, H$) plane for the case $V\leq0$. From this figure, one can see that  $a_{Es+}$ represents the center point  and the universe can stay at the stable state eternally. If $a$ initiates from close to the center point, the universe will undergo an infinite oscillation represented by circles around $a_{Es+}$.

\begin{figure}[htbp]
\includegraphics[width=7cm]{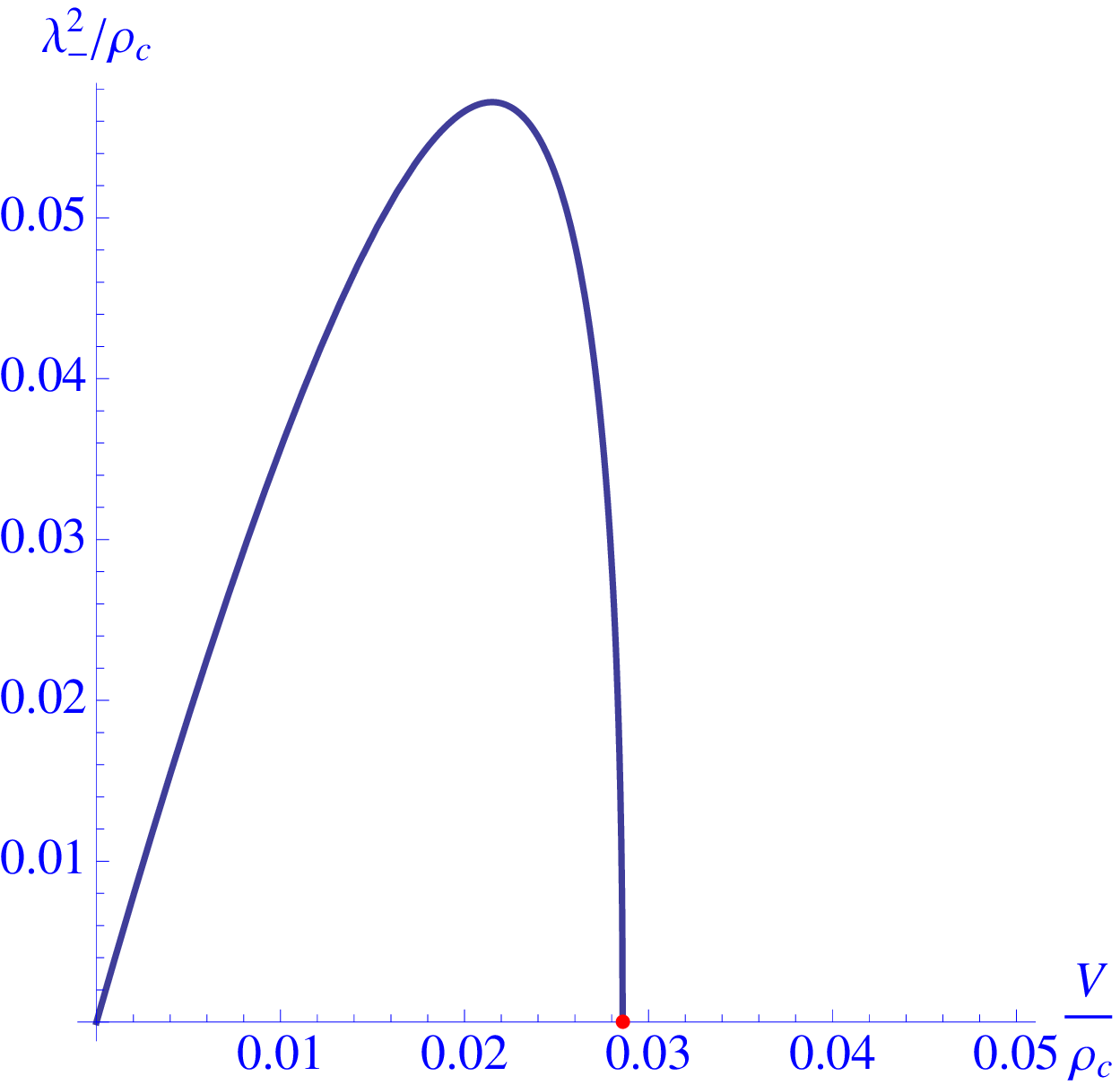}\quad\includegraphics[width=7cm]{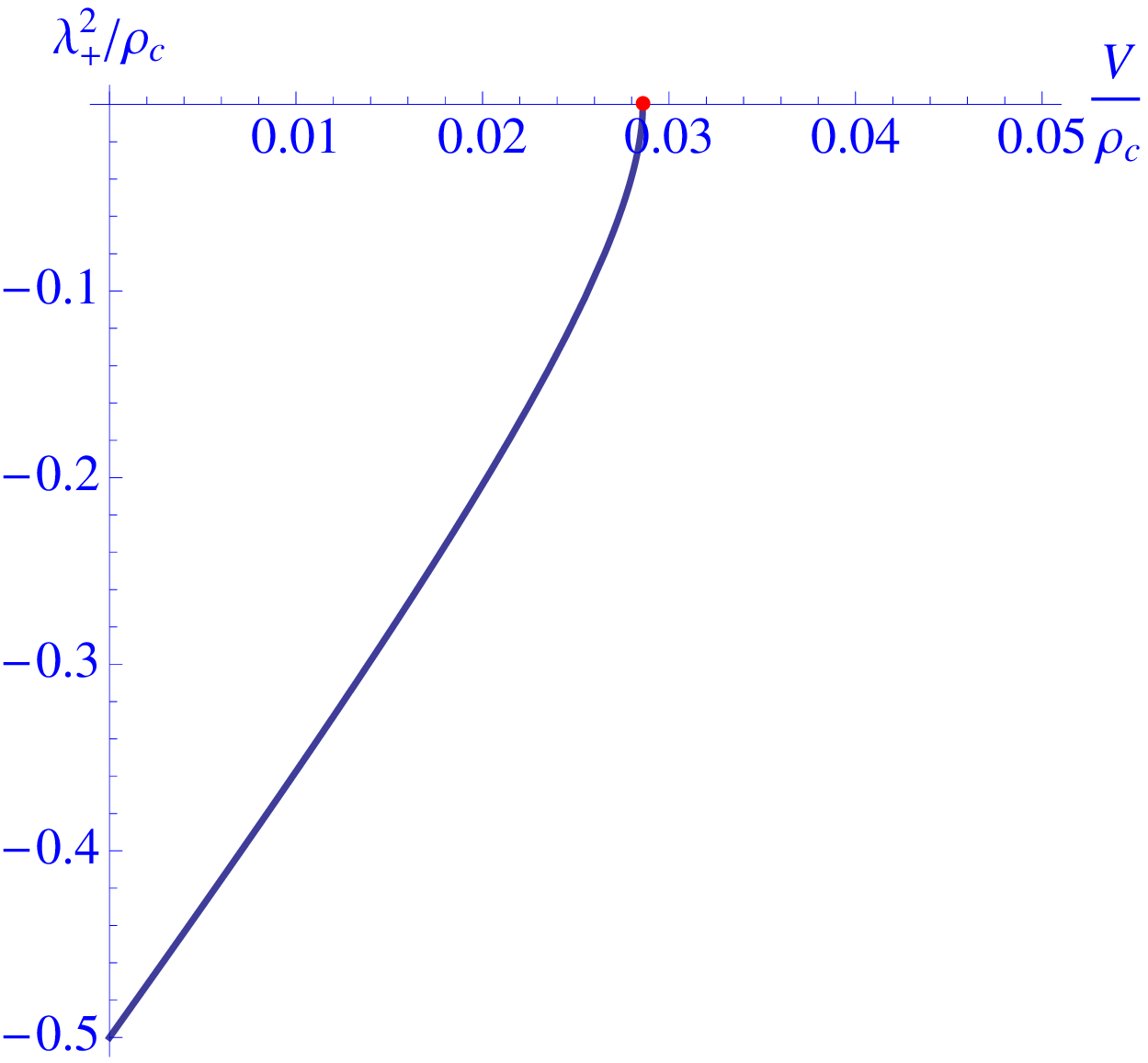}
\caption{\label{f2} The eigenvalues $\lambda^2_+$ and $\lambda^2_-$ for $0<V\leq V_{crit}$. Red point corresponds to $V=V_{crit}$.  }
\end{figure}

\begin{figure}[htbp]
\includegraphics[width=7cm]{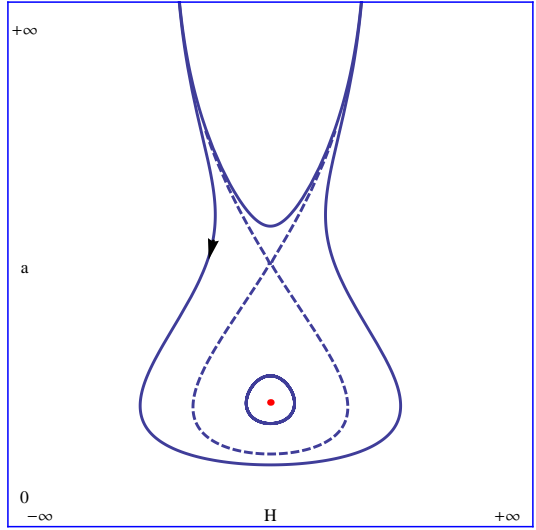}
\caption{\label{f3} The phase diagram in  ($a, H$) plane for $0<V< V_{crit}$.  The parameters are set as $\rho_c=1$, $C=-1$,  and $V=0.018$. The red solid point
denotes the stable center point.
The dotted curve represents a separatrix of stable region and unstable one, and the saddle point occurs at the point where the separatrix self-intersects. Different curves correspond to the cases of $a$ with different initial values.   The arrow represents the direction of time evolution. The axes have been compactified using the relations $x(t) = \arctan(H)$ and $y(t) = \arctan(\ln a)$. } %The axes have been compactified using the relations $x = \arctan(H)$ and  $y= \arctan(\ln \dot{\phi})$.}
\end{figure}

\begin{figure}[htbp]
\includegraphics[width=7cm]{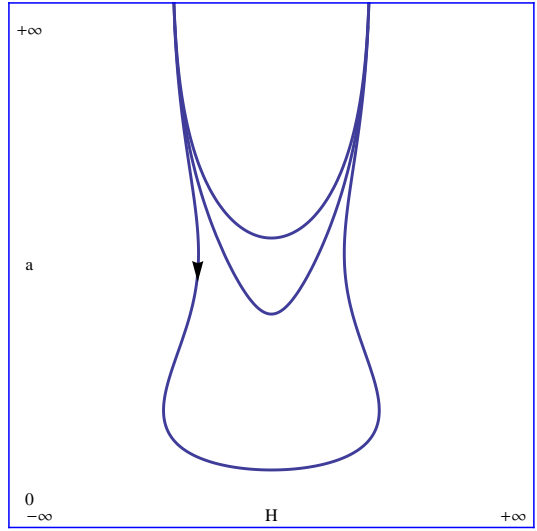}
\caption{\label{f23} The phase diagram in  (${a, H}$) plane for $V>V_{crit}$.  The parameters are set as $\rho_c=1$, $C=-1$ and $V=0.03$. There is no stable static state solution.  A bounce scenario is obtained.  The arrow represents the direction of time evolution. The axes have been compactified using the relations $x(t) = \arctan(H)$ and $y(t) = \arctan(\ln a)$. }
\end{figure}

If $0<V<V_{crit}$, there are two Einstein static state  solutions: $a_{Es}=a_{Es+}$ and  $a_{Es}=a_{Es-}$,  which coincide each other and  become unstable when $V=V_{crit}$. The eigenvalues of  $a_{Es+}$ and  $a_{Es-}$ are $\lambda^2_+$ and $\lambda^2_-$, respectively, which are shown in Fig.~(\ref{f2}). We find that $\lambda^2_+<0$ and $\lambda^2_->0$.  Thus, $a_{Es}=a_{Es+}$ is a stable center point, while $a_{Es}=a_{Es-}$ is a saddle one. Fig.~(\ref{f3}) gives the  phase portraits  in ($a, H$) plane for the case of $0<V<V_{crit}$ with different initial values of $a$. The solid point
denotes the stable center point.
The dotted curve represents a separatrix of stable region and unstable one, and the saddle point occurs at the point where the separatrix self-intersects.  If $a$ is close to the stable  center point initially,  the universe may undergo an infinite oscillation, which is shown as the solid circle  in Fig.~(\ref{f3}). If trajectories
pass above the saddle point, the universe collapses and then bounces into an  inflationary era.
If $V>V_{crit}$, there is no stable critical point. As shown in Fig.~(\ref{f23}), only a bounce scenario is obtained.

\begin{figure}[htbp]
\includegraphics[width=9cm]{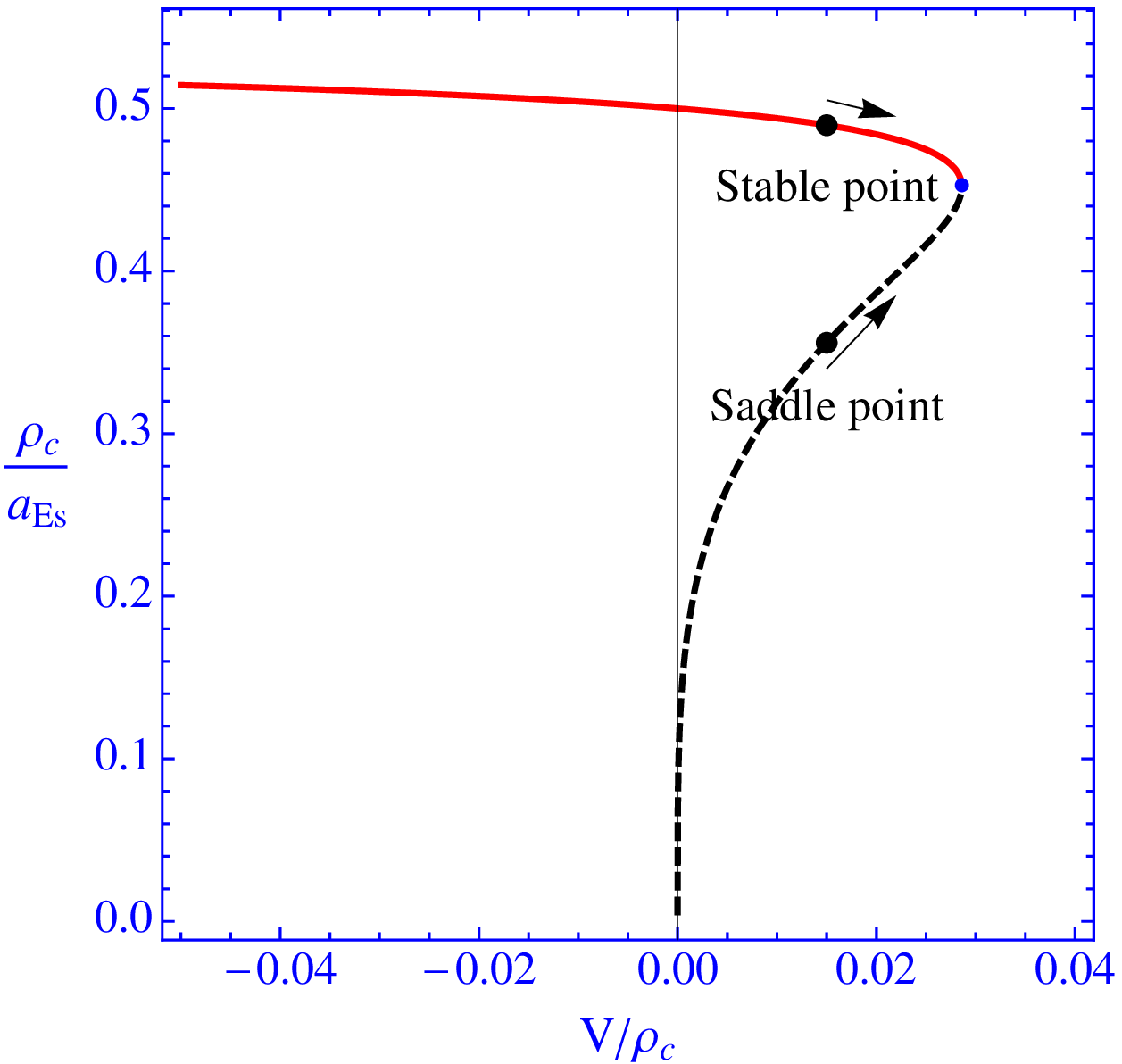}
\caption{\label{f4} The evolutions of the stable point and unstable one  with the scalar field climbing up its potential slowly.
 }
\end{figure}

 For the purpose of avoiding the big bang singularity  using the emergent scenario, we assume that $V\rightarrow constant<V_{crit}$ when $t\rightarrow -\infty$ and $a$ is close to the stable center point initially. Thus, the universe can stay at the stable state past eternally.  In order to render  the universe evolve from this stable  state to an inflation phase naturally, we further require that the potential increases very slowly from the infinite past so that  it reaches $V=V_{crit}$ at a  certain point as time goes on. As shown in Fig.~(\ref{f4}), with the scalar field climbing up its potential slowly the stable point moves closer and closer to the saddle one.  Once the potential reaches  the critical value $V_{crit}$, the stable critical point coincides with the saddle one and it becomes unstable.  Then the universe enters  an inflationary epoch.

\section{An emergent potential}

\begin{figure}[htbp]
\includegraphics[width=9cm]{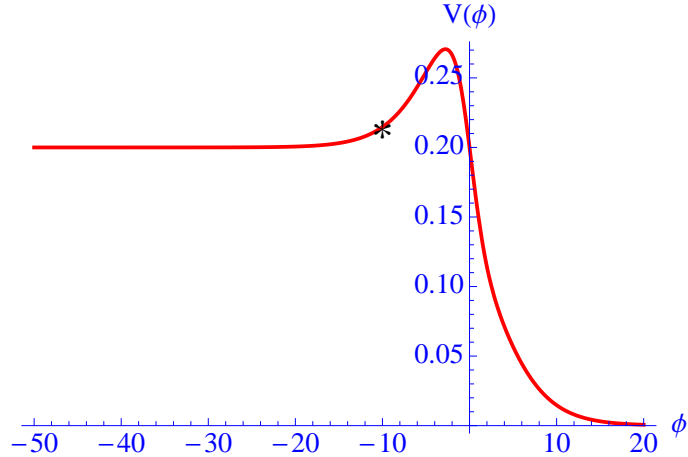}
\caption{\label{fv} The evolution of the potential given in Eq.~(\ref{fv21}) with  model parameters being $V_0=0.1$, $V_1=0.2$, $c_1=2$ and $c_2=3.1$. Black  star represents the critical value $V_{crit}$.
 }
\end{figure}

In this  section, we propose a potential of the scalar field which successfully realizes a spatially flat emergent universe, i.e., a potential that avoids the big bang
singularity and naturally leads to inflation,
\begin{eqnarray}\label{fv21}
V(\phi)=V_0 \bigg (1+\tanh\big(- {\phi \over c_1}\big) \bigg)+V_1 \frac{1}{1+\cosh({\phi \over c_2})}\;,
\end{eqnarray}
where $V_0$, $V_1$,  $c_1$ and $c_2$ are constants.
In Fig.~(\ref{fv}), we plot the evolutionary curve of this potential with model parameters being $V_0=0.1$, $V_1=0.2$, $c_1=2$ and $c_2=3.1$.   This potential is completely different from  the one given in~\cite{Ellis20041} for a spatially closed emergent universe. This is because in the spatially flat SS braneworld  the potential is required to be a constant in the asymptotic past and then increase slowly to ensure that the universe exits naturally  from an Einstein static state, rather than drops slowly from its original Einstein static value as that is required in the spatially closed case considered in~\cite{Ellis20041}.

\begin{figure}[htbp]
\includegraphics[width=7cm]{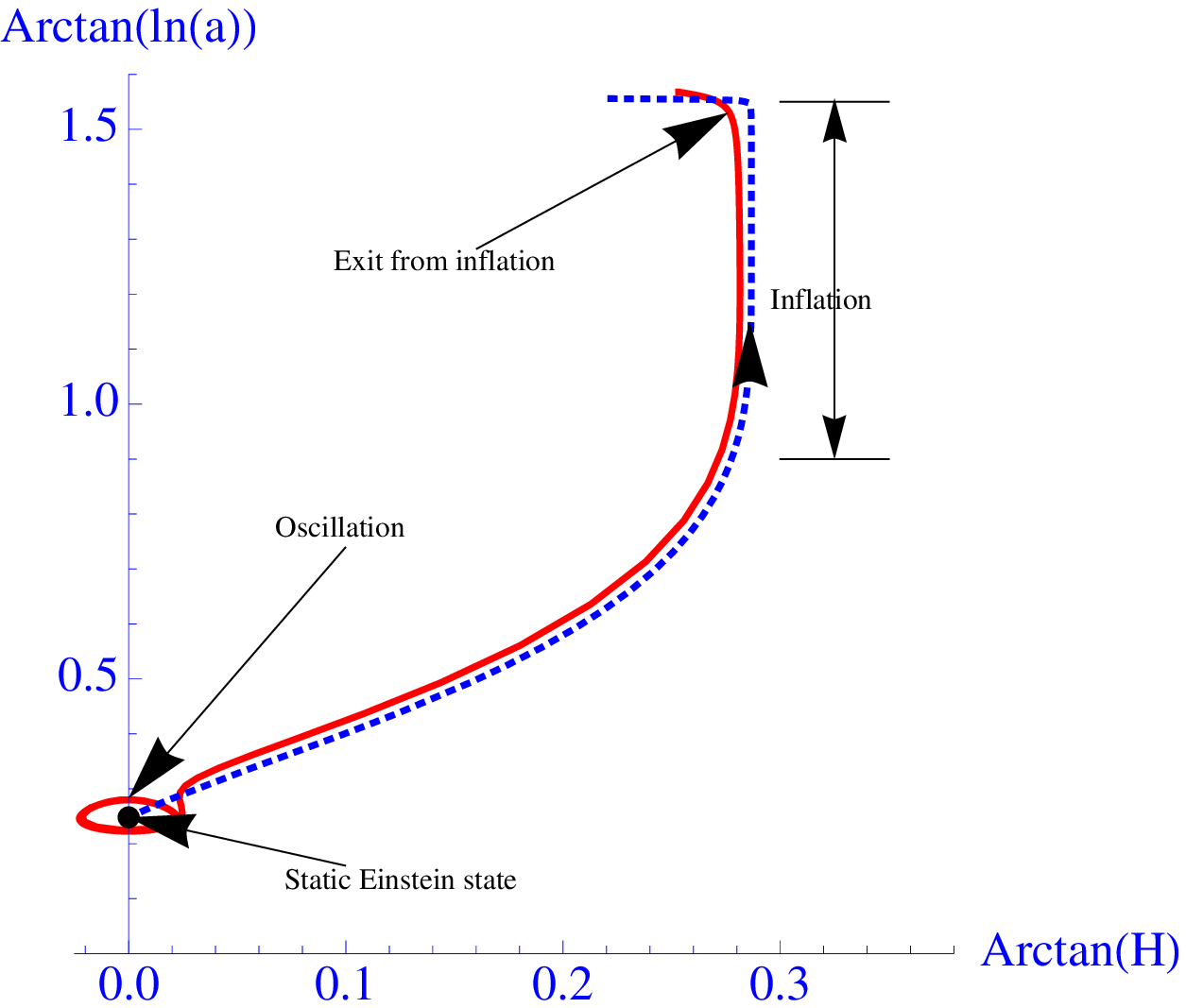}\includegraphics[width=7cm]{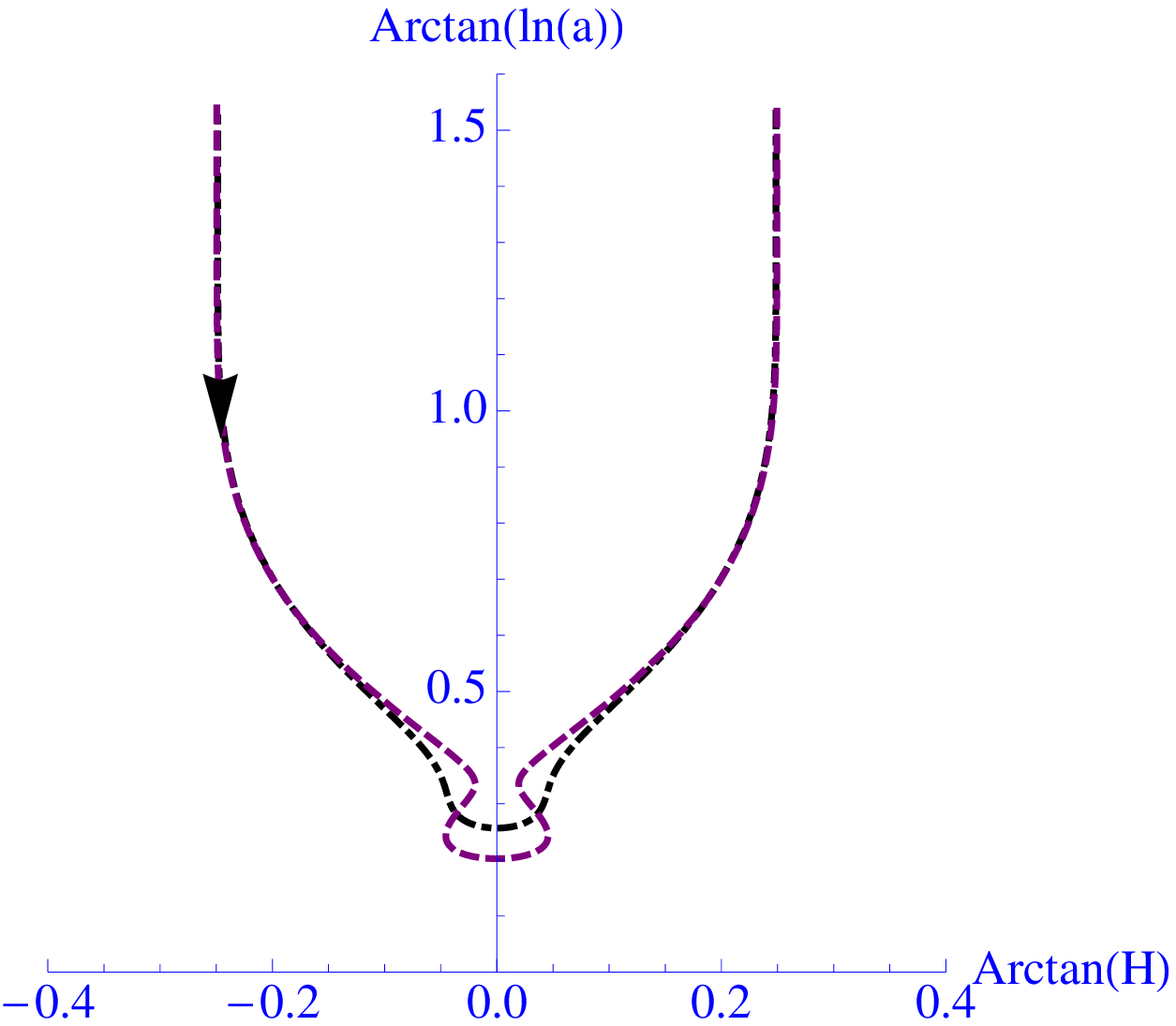}
\caption{\label{fa} The phase portraits in ($a, H$) plane with the potential given in Eq.~(\ref{fv21}). The arrow represents the direction of time evolution. The blue dotted and red solid lines in the left panel correspond to the cases with  the initial value of $a$ coinciding  with and being close to the stable center point, respectively.   In the right panel the purple dashed and black dot-dashed   lines correspond to the cases of   the initial value of $a$ significantly deviating   from  the stable center point and $V>V_{crit}$, respectively.     %The evolutionary curves of the cosmic scale factor,  the  Hubble parameter  and the potential given in Eq.~(\ref{fv21}) with time under the condition that $a$ is equivalent  with the stable center point initially.
 The  model parameters are set as  $V_0=0.1$,  $V_1=0.2$,  $c_1=2$,  $c_2=3.1$, $C=-1$ and $\rho_c=7.5$ except for the black dot-dashed line, where $\rho_c=6$ is chosen. }
\end{figure}

 Substituting this potential into the scalar field equation and Eq.~(\ref{ddota/a}), and solving  numerically these two equations, we can get the evolutionary behaviors of the cosmic scale factor.
 % and the scalar field $\phi$.
 In Fig.~(\ref{fa}), we show our numerical  results  with different initial values of  $a$.  As expected,  %in the infinite past ($t\rightarrow -\infty$ or $\phi \rightarrow -\infty$) the potential approaches a constant, the scalar field rolls on it with a constant speed.
  if $a$ coincides with  $a_{Es+}$ given in Eq.~(\ref{aes}) initially,   the universe stays at the stable  Einstein static state ($a=a_{Es+}$) past eternally, whereas if  $a$ is close to $a_{Es+}$  initially,   the universe undergoes an infinite oscillation. These are shown in the left panel of Fig.~(\ref{fa}) as the dotted and solid lines, respectively.  With the scalar field climbing up its potential, the stability condition ($V<V_{crit}$) of  Einstein static state solution is broken and the universe exits from its stable state. When the scalar field rolls over the maximum value of its potential, it enters a slow-roll era and the universe begins to inflate.   As shown in the left panel of  Fig.~(\ref{fa}),  after a long enough period of exponential expansion the universe can exit gracefully from inflation.


However,  if $V>V_{crit}$ or the initial value of $a$  significantly deviates  from  the stable center point, as those are shown as the dot-dashed and  dashed lines in the right panel of Fig.~(\ref{fa}), then a bouncing universe  is obtained in accordance  with the results given in Figs.~(\ref{f3}, \ref{f23}).

%\begin{figure}[htbp]
%\includegraphics[width=7cm]{figa2}
%\caption{\label{fa2} The evolutionary curve of  the cosmic factor  with time under the condition that  the initial value of $a$   is outside the stable region shown in Fig.~(\ref{f3}). The  model parameters are set as  $V_0=0.1$, $V_1=0.2$, $c_1=2$, $c_2=3.1$, $C=-1$ and $\rho_c=7.5$.  }
%\end{figure}

\section{Conclusions}
 Emergent scenario suggests that our universe originates from an Einstein static state universe rather than a big bang singularity.  It thus provides a way to resolve the big bang singularity problem. However, the original idea of an emergent universe is not successfully implemented in classical general relativity,  since there is no  stable Einstein static  state solution. With the effect of  quantum gravity, extra dimension, modified gravity and so on taken into consideration, it has been found a stable Einstein static state  universe can be obtained in both positive and negative spatial curvature cases. However, the present cosmological observations seem to favor a spatially flat universe.
So, in this paper, we study the stability of Einstein static sate solutions in  a spatially flat universe. The SS  braneworld scenario with a negative dark radiation term is considered in our discussion.  We find that, when the potential of the scalar field, which is assumed to be the only matter energy component, is negative or zero, there is a stable Einstein static state  solution.   For the case $0<V<V_{crit}$, there are two critical solutions.  One is  a center equilibrium point, and the other is a saddle one.  Therefore, when $V<V_{crit}$,  the universe can stay at a stable state  eternally. If the scale factor $a$ is close to the stable center point initially, the universe may undergo an infinite oscillation until the potential reaches  a critical value  $V_{crit}$. When $V=V_{crit}$,  the stable center point coincides with the saddle one and becomes unstable. As a result, the universe can exit the Einstein static state  as the scalar field climbs up its potential slowly.

We also propose an example of such a potential that successfully implements an emergent universe in a spatially flat universe. However, we must point out that here we only discuss the stability of  Einstein static state solution against the
homogeneous perturbations and whether the universe can enter the inflation era naturally with the potential we propose. Extending our stability analysis to the case of the inhomogeneous perturbations as well as studying
other aspects of the potential, such as the power spectrum generated and so on,  are interesting further topics and will be left for  future investigations.

\acknowledgments  This work was supported by
the National Natural Science Foundation of China under Grants Nos.
Nos. 10935013, 11075083, 11175093, 11222545 and
11375092, Zhejiang Provincial Natural Science
Foundation of China under Grants Nos. Z6100077 and R6110518, the
FANEDD under Grant No. 200922, the National Basic Research Program
of China under Grant No. 2010CB832803,  the SRFDP under Grant No. 20124306110001,  the PCSIRT under Grant No. IRT0964 and the Hunan Provincial
Natural Science Foundation of China under Grant No. 11JJ7001.

%%%%%%%%%%%%%%%%%%%%%%%%%%%%%%%%%%


\begin{thebibliography}{99}
\bibitem{Starobinsky1980} A. Starobinsky, Phys. Lett. B {\bf91}, 99 (1980).
\bibitem{Guth1981} A. Guth, Phys. Rev. D {\bf23}, 347 (1981).
\bibitem{Ellis20041} G. F. R. Ellis, R. Maartens, Class. Quant. Grav. {\bf21},  223 (2004).
\bibitem{Ellis20042} G. F. R. Ellis, J. Murugan, and C. G. Tsagas, Class. Quant. Grav. {\bf21}, 233 (2004).
\bibitem{Eddington} A. S. Eddington, Mon. Not. Roy. Astron. Soc. {\bf 90}, 668 (1930).
\bibitem{Gibbons} G. W. Gibbons, Nuc. Phys. B {\bf 292}, 784 (1987).
\bibitem{Barrow0} J. D. Barrow, G. F. R. Ellis, R. Maartens and C. G. Tsagas, Class. Quant. Grav. {\bf 20}, L155 (2003).
\bibitem{Gergely} L. A. Gergely and R. Maartens, Class. Quant. Grav. {\bf 19}, 213 (2002);
                            A. Gruppuso, E. Roessl and M. Shaposhnikov, JHEP {\bf 011},  0408 (2004);
                            S. S. Seahra, C. Clarkson and R. Maartens, Class. Quant. Grav. {\bf 22}, L91 (2005);
                            S. Carneiro, R. Tavakol, Phys. Rev. D {\bf 80},  043528 (2009);
                            A. Odrzywolek, Phys. Rev. D {\bf 80},  103515 (2009);
                            C. Clarkson and S. S. Seahra, Class. Quant. Grav. {\bf 22}, 3653 (2005);
                            C. G. Bohmer, Class. Quant. Grav. {\bf 21}, 1119 (2004);
                            C. G. Bohmer, L. Hollenstein and F. S. N. Lobo, Phys. Rev. D {\bf 76}, 084005 (2007);
                            R. Goswami, N. Goheer and P. K. S. Dunsby, Phys. Rev. D {\bf 78}, 044011 (2008);
                            N. Goheer, R. Goswami and P. K. S. Dunsby, Class. Quant. Grav. {\bf 26}, 105003 (2009);
                            S. S. Seahra and C. G. Bohmer, Phys. Rev. D {\bf 79}, 064009 (2009);
                            C. G. Bohmer and F. S. N. Lobo, Phys. Rev. D {\bf 79}, 067504 (2009);
                            C. G. Bohmer, L. Hollenstein, F. S. N. Lobo and S. S. Seahra, The Twelfth Marcel Grossmann Meeting {\bf 379}, 1977 (2012);  %, chap 379, p. 1977-1979 arXiv:1001.1266 [gr-qc];
                            K. Zhang, P. Wu and H.  Yu, Phys. Lett. B {\bf 690},  229 (2010);
                            L. Parisi, M. Bruni, R. Maartens and K. Vandersloot, Class. Quant. Grav. {\bf24}, 6243 (2007);
                            M. i. Park, JHEP {\bf 0909},  123 (2009);
                            P. Wu, H. Yu, J. Cosmol. Astropart. Phys. {\bf 0905},  007  (2009);
                            K. Maeda, Y. Misonoh, T. Kobayashi,  Phys. Rev. D {\bf 82},  064024 (2010);
                            C. G. Bohmer and F. S. N. Lobo, Eur. Phys. J. C {\bf 70}, 1111 (2010);
                             P. Wu and H.  Yu, Phys. Rev. D {\bf 81}, 103522 (2010);
                             A. T. Mithani, A. Vilenkin,  arXiv:1110.4096;
                             A. T. Mithani, A. Vilenkin,  arXiv:1204.4658;
                             K. Zhang, P. Wu and H.  Yu, Phys. Rev. D.  {\bf 85}, 043521 (2012);
                            J. T. Li, C. C. Lee and C. Q. Geng,  Eur. Phys. J. C {\bf 73}, 2315 (2013);
                            L. Parisi, N. Radicella, G. Vilasi, Phys. Rev. D {\bf 86}, 024035 (2012);
                              K. Zhang, P. Wu and H.  Yu,  Phys. Rev. D. {\bf 87},  063513 (2013);
                              C. G. Boehmer, F. S. N. Lobo, N. Tamanini, arXiv:1305.0025;
                              A. Vilenkin,  arXiv:1305.3836;
                              A. Aguirre, J. Kehayias,  Phys. Rev. D {\bf 88}, 103504 (2013).

\bibitem{Barrow}%J. D Barrow, G. Ellis, R. Maartens, C. Tsagas, Class. Quant. Grav. {\bf 20},  L155 (2003);
                                T. Clifton and J. D. Barrow, Phys. Rev. D {\bf 72}, 123003 (2005);
                                D. J. Mulryne, R. Tavakol, J. E. Lidsey and G. F. R. Ellis, Phys. Rev. D {\bf 71}, 123512 (2005);
                                J. D. Barrow, C. G. Tsagas, Class. Quant. Grav. {\bf 26}, 195003 (2009).
\bibitem{Canonico} R. Canonico and L. Parisi, Phys. Rev. D {\bf 82}, 064005
(2010).
\bibitem{Wu}  P. Wu and H. Yu, Phys. Lett. B {\bf 703}, 223 (2011).
\bibitem{Parisi}L. Parisi, N. Radicella, and G. Vilasi, Phys. Rev. D {\bf 86}, 024035 (2012).
\bibitem{Mulryne} J. E. Lidsey, D. J. Mulryne, Phys. Rev. D {\bf 73},  083508  (2006).
\bibitem{Komatsu2011}   E. Komatsu, et al., Astrophys. J. Suppl. {\bf 192}, 18  (2011).
\bibitem{Shtanov2003} Y. Shtanov, V. Sahni, Phys. Lett. B {\bf 557}, 1 (2003).
\bibitem{Randall19991}  L. Randall, R. Sundrum, Phys. Rev. Lett. {\bf 83}, 3370 (1999).
\bibitem{Randall19992}  L. Randall, R. Sundrum, Phys. Rev. Lett. {\bf 83}, 4690 (1999).
\bibitem{Yndurain} F. J. Yndurain, Phys. Lett. B {\bf 256}, 15 (1991).
\bibitem{Dvali}G. Dvali, G. Gabadadze and G. Senjanovic, hep-ph/9910207.
\bibitem{Ya}Ya. Arefeva, B. G. Dragovic and I. V. Volovich, Phys. Lett. B {\bf 177}, 357 (1986).
\bibitem{Chaichian} M. Chaichian and A. B. Kobakhidze, Phys. Lett. B {\bf 488}, 117 (2000).
\bibitem{Maier} R. Maier, N. Pinto-Neto, I. D. Soares, Phys. Rev. D {\bf 87}, 043528 (2013).
\bibitem{Iglesias} A. Iglesias and Z. Kakushadze, Phys. Lett. B {\bf 515}, 477 (2001).
\bibitem{Ya2} Ya. Arefeva and I. V. Volovich, Phys. Lett. B {\bf 164}, 287 (1985).
\bibitem{Shiromizu}T. Shiromizu, K. Maeda, and M. Sasaki, Phys. Rev. D {\bf 62}, 024012 (2000).
\bibitem{Mukohyama} S. Mukohyama, Phys. Lett. B {\bf 473}, 241 (2000).
\bibitem{Ida} D. Ida, JHEP {\bf 0009}, 014 (2000).
\bibitem{Ashtekar}A. Ashtekar, P. Singh, Class. Quant. Grav. {\bf 28},  213001 (2011).
\bibitem{Markov}  M. A. Markov and  V. F. Mukhanov, JETP Lett. {\bf 40}, 1043 (1984); Pisma Zh.Eksp.Teor.Fiz. {\bf 40}, 265 (1984).




\end{thebibliography}
\end{document}